\documentclass[12pt,epsfig]{article}
\usepackage{graphicx}
\font\tenrm=cmr10

 1
\font\elevenrm=cmr10 scaled\magstep 1

\renewenvironment{thebibliography}[1]
 { \elevenrm
   \begin{list}{\arabic{enumi}.}
    {\usecounter{enumi}     \setlength{\parsep}{0pt}
     \setlength{\itemsep}{3pt} \settowidth{\labelwidth}{#1.}
     \sloppy
    }}{\end{list}}

\begin{document}
\title{\Large "Buddha's light" of cumulative particles
\footnote{Presented at the International Seminar
Quarks-2014, Suzdal',Russia, 2-8 June 2014}}

\author{V.B.~Kopeliovich$^{a,b}$\footnote{{\bf e-mail}: kopelio@inr.ru}, 
G.K.~Matushko$^a$\footnote{{\bf e-mail}: matushko@inr.ru} and I.K.~Potashnikova$^c$\footnote{{\bf e-mail}: irina.potashnikova@usm.cl} 
\\
\small{\em a) Institute for Nuclear Research of RAS, Moscow 117312, Russia} \\
\small{\em b) Moscow Institute of Physics and Technology (MIPT), Dolgoprudny, 
Moscow district, Russia} \\
\small{\em c) 
Departamento de F\'{\i}sica, Universidad T\'ecnica Federico Santa Mar\'{\i}a;}\\
\small{\it and Centro Cient\'ifico-Tecnol\'ogico de Valpara\'iso,
Avda. Espa\~na 1680, Valpara\'iso, Chile}}
\maketitle
{\rightskip=2pc
 \leftskip=2pc
 \noindent}
{\rightskip=2pc
 \leftskip=2pc

To the memory of Lyonya Kondratyuk, outstanding scientist and person

\tenrm\baselineskip=11pt
\begin{abstract}
{We show analytically that in the cumulative particles production off
nuclei multiple interactions lead to 
a glory-like backward focusing effect. Employing 
the small phase space method we arrived at a characteristic angular dependence of the production cross section
$d\sigma \sim 1/ \sqrt {\pi - \theta}$ near the strictly backward direction. This effect takes place 
for any number $n\geq 3 $ of interactions of rescattered particle, either elastic or inelastic (with  resonance excitations in 
intermediate states), when the final particle is produced near corresponding kinematical boundary.
In the final angles interval including the value $\theta =\pi$ the angular dependence of the
cumulative production cross section can have the crater-like (or funnel-like) form.
Such a behaviour of the cross section near the backward direction is in qualitative agreement with some
of available data. Explanation of this effect and the angular dependence of the cross section near $\theta \sim \pi$
are presented for the first time.} \end{abstract}
 \noindent
\vglue 0.3cm}
\newpage 
\section{Introduction} 
Intensive studies of the particles production processes in high energy interactions
of different projectiles with nuclei, in regions forbidden by kinematics for the 
interaction with a single free nucleon, began back in the  70th mostly at JINR  (Dubna) and ITEP (Moscow).
 Relatively simple experiments could provide 
information about such objects as fluctuations of the nucleus density \cite{blokh} or, discussed much
later, few nucleon (or multiquark) clusters probably
existing in nuclei. At JINR such processes have been called "cumulative production"
\cite{baldin1,baldin2}, at ITEP the variety of properties of such reactions has been called
"nuclear scaling" \cite{leksin1}- \cite{leksin3} because certain universality of these properties
has been noted, confirmed somewhat later at much higher energy,
$400\, GeV$ incident protons \cite{fran-le-1,fran-le-2} and $40\,GeV/c$ incident pions, kaons and antiprotons
\cite{antip1, antip2}.
A new wave of interest to this exciting topic appeared lately. New experiment has been performed
in ITEP \cite{abramov} aimed to define the weight of multiquark configurations in the carbon
nucleus \footnote{We do not pretend here to give a comprehensive review of numerous experiments on
cumulative particles production.}.

The interpretation of these phenomena as being manifestation of internal structure of
nuclei assumes that the secondary interactions, or, more generally, multiple
interactions processes (MIP) do not play a crucial role in such production \cite{frastri1} - \cite{galoian}. 
Generally, the role of secondary interactions in the particles production off nuclei 
is at least two-fold: they decrease the amount
of produced particles in the regions, where it was large (it is, in particular, the screening phenomenon), 
and increase the production probability in regions where it was small;
so, they smash out the whole production picture.

The development of the Glauber approach \cite{glauber,glauber2} to the description of particles scattering off
nuclei has been considered many years ago as remarkable progress in understanding
the particles-nuclei interactions. Within the Glauber model the amplitude of the                                 
particle-nucleus scattering is presented in terms of elementary particle-nucleons
amplitudes and the nucleus wave function describing the nucleons distribution inside
the nucleus. The Glauber screening correction for the total cross section of
 particle scattering
off deuteron allows widely accepted, remarkably simple and transparent interpretation.

Gribov \cite{gribov} explained nontrivial peculiarities of the space-time picture of such scattering
processes and concluded that the inelastic shadowing corrections play an important role at
high enough energy and should be included into consideration. \footnote{The pion double charge exchange scattering 
is an interesting example of the reaction where the inelastic intermadiate states give the dominant 
contribution at high enough energy \cite{kaikru,kruten}.} 

In the case of the large angle particle production the background processes which mask 
the possible manifestations of nontrivial details of nuclear structure, are 
subsequent multiple interactions with nucleons inside the nucleus leading to the
particles emission in the "kinematically forbidden" region.
Leonid Kondratyuk was the first who the has noted that rescattering
of intermediate particles could lead to the final particles emission in
"kinematically forbidden" regions (KFR).
The rigorous investigation of the double interaction process in the case of pion
production off deuteron (see Fig. 1.1) has been made first by L.Kondratyuk and V.Kopeliovich in
\cite{konkop}. Later the multiple interaction processes leading to nucleons
production in KFR were investigated in \cite{kop1} and in more details in \cite{kop2}
where the magnitude of the cumulative protons production cross sections was estimated as well.

\vglue 0.1cm
\begin{figure}[h]
\begin{center}

\includegraphics[natwidth=800,natheight=600,width=7.cm,angle=0]{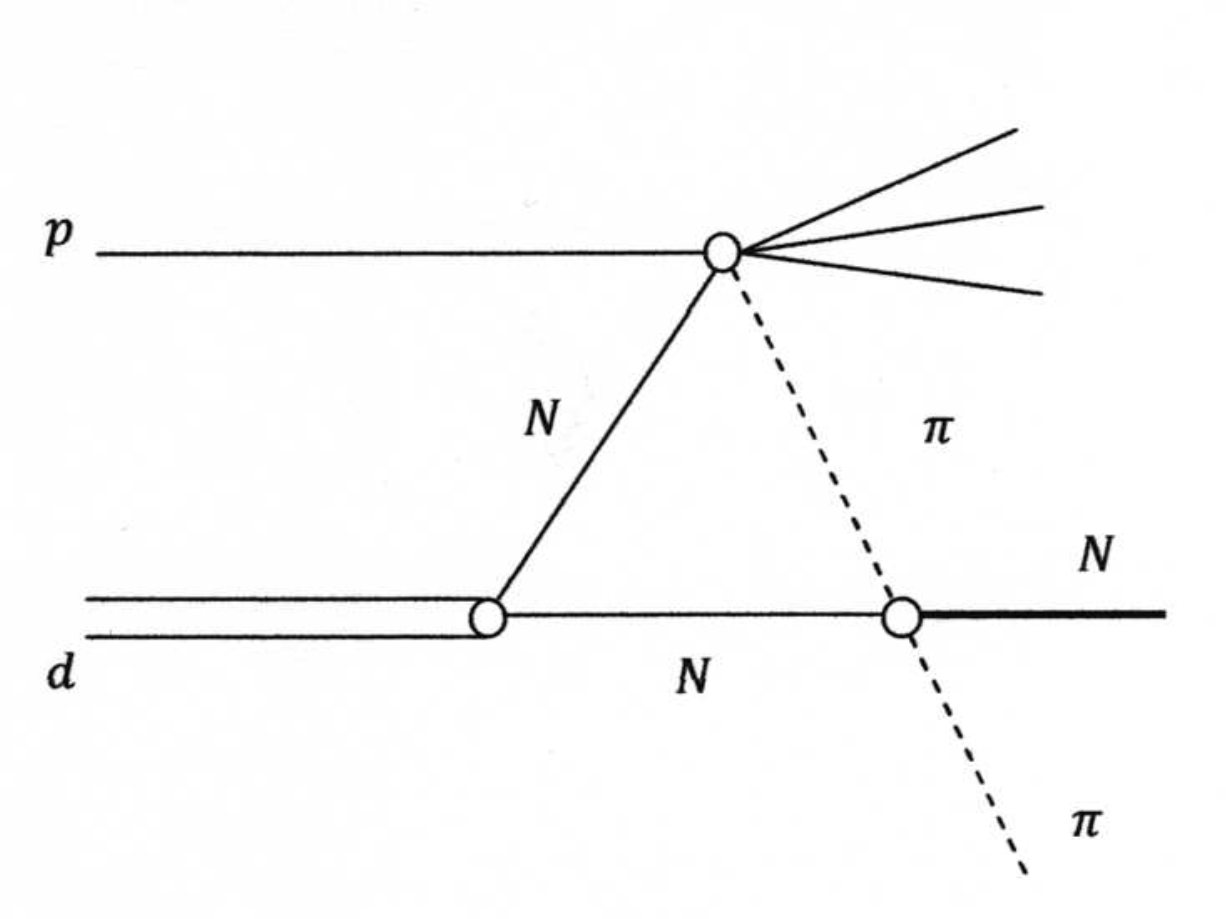}
\end{center}
{\small{\bf Fig.\ 1.1.} The simplest pion rescattering diagram which leads to the partial
fill-up of the "kinematically forbidden" region for the case of the cumulative pion production
by protons on the deuteron \cite{konkop}. }
\end{figure}
\vglue 0.1cm

 M.A.Braun and V.V.Vechernin with coauthors  made many interesting and important observations and investigated
processes leading to the particles emission in KFR \cite{bv-res-1}-\cite{braun-lepto}, including the processes 
with resonances in intermediate state \cite{bv-res-1}-\cite{bv-res-2}.
They found also that processes with pions in intermediate state
lead to the nucleons emission in KFR due to subsequent processes, like $\pi\,N \to N\,\pi$ \cite{braun-pin1,braun-pin2}.
Basic theoretical aspects of MIP leading to the cumulative particles emission and some review of the situation 
in this field up to 1985 have been presented in \cite{long}.

Several authors attempted the cascade calculations of cumulative particles production
cross sections relying upon the available computing codes created previously 
\cite{sibste} - \cite{nomad2}.
The particles production cross section was found to be in reasonable agreement with data.
Different kinds of subprocesses play a role in these calculations, and certain work
should be performed for detailed comparison. In calculations by NOMAD Collaboration the particles
formation time has been considered as a parameter, and results near to the experimental
observations have been obtained for this time equal to $\sim 2 \,Fm$ \cite{nomad1,nomad2}, see discussion below.

While many authors have admitted the important role of the final state interactions (FSI), most of them 
did not discuss the active role of such interactions, i.e. their contribution to particles production 
in KFR, see e.g. \cite{size}. It has been stated in a 
number of papers that multiple interactions cannot describe the spectra of backwards
emitted particles. Such statement in fact has no firm grounds because there were so far no reliable
calculations of the MIP contributions to the cross sections and other
observables in the cumulative particles production reactions.  Moreover, such calculations are hardly possible 
because, as we argue in the present paper,  necessary
information about elementary interactions amplitudes is still lacking.

Several specific features of the MIP mechanism have been noted previously experimentally and 
discussed theoretically \cite{kop2,kopmix, long},
among them the presence of the recoil nucleons, which amount grows with increasing energy of the
cumulative particle, possible large value of the cumulative baryons polarization, and some other, 
see \cite{long}. The enhancement of the production cross section near the strictly backward
direction has been detected in a number of experiments, first at JINR (Dubna) \cite{baldin-77, foc-1} and somewhat later at
ITEP (Moscow) \cite{foc-2, foc-4}. This glory-like effect
which can be called also the "Buddha's light" of cumulative particles, has been shortly discussed previously
in \cite{kop2, long}. More experimental evidence of this effect appeared since that time \cite{glo-1, glo-2}. 
 Here we show analytically that presence of the backward focusing  effect
is an intrinsic property of the multiple interaction mechanism leading to the cumulative particles production.
The detailed treatment of this effect is presented, including the angular dependence of the particles production
cross section near the strictly backward direction. To our knowledge, the proof of the existence of the nuclear glory 
phenomenon was absent so far in the literature. 

In the next section the peculiarities of kinematics of the processes in KFR will be recalled,
in section 3 the small phase space method of the MIP contributions calculation to the particles
production cross section in KFR is described. In section 4 the focusing effect, similar to the known in optics glory
phenomenon, is described in details.
Final section contains discussion of problems and conclusions.
Some mathematical aspects of the nuclear glory phenomenon are presented in Appendix.

\section{Details of kinematics}
When the particle with 4-momentum $p_0=(E_0, \vec p_0)$ interacts with the nucleus with the mass $m_t \simeq A m_N$,
and the final particle of interest has the 4-momentum $k_f=(\omega_f, \vec k_f)$ the basic kinematical relation is
$$ (p_0+ p_t - k_f)^2 \geq M_f^2, \eqno (2.1) $$
where $M_f$ is the sum of the final particles masses, except the detected particle of interest.
At large enough incident energy, $E_0 \gg M_f$, we obtain easily
$$ \omega_f - z k_f \leq m_t, \eqno (2.2) $$
which is the basic restriction for such processes. $z=cos\,\theta < 0$ for particle produced
in backward hemisphere. The quantity $(\omega_f - z k_f)/m_N$ is called the cumulative number
(more precize, the integer part of this ratio plus one).

\vglue 0.1cm
\begin{figure}[h]
\begin{center}
\includegraphics[natwidth=800,natheight=600,width=9.cm,angle=0]{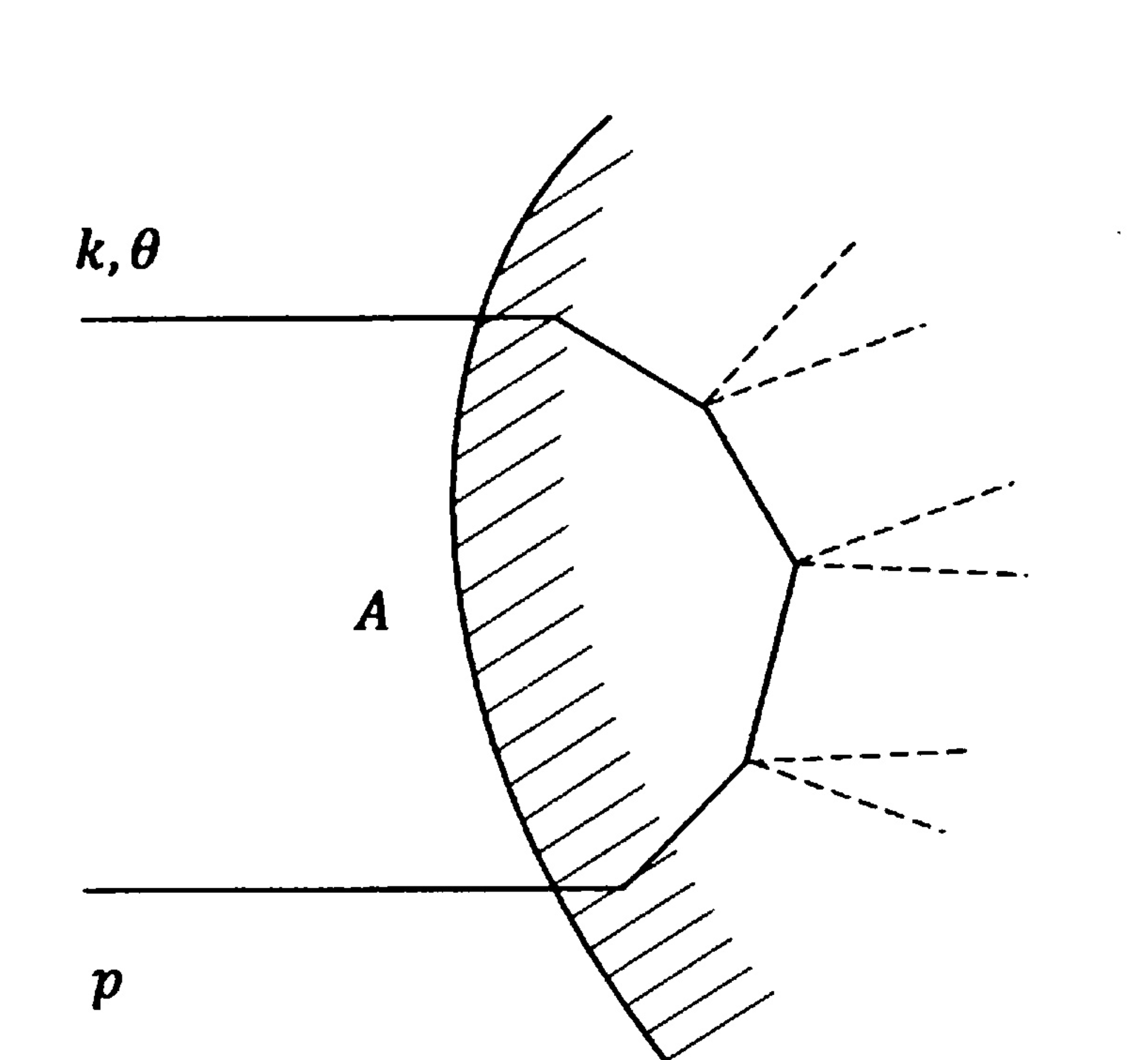}

\end{center}
{\small {\bf Fig.\ 2.1.} Schematical picture of the multiple interaction process within the nucleus $A$
leading to the emission of the final particle with the momentum $k$ at the angle $\theta$ relative
to the projectile proton momentum. The binary reactions are assumed to take place in secondary interactions.}
\end{figure}
\vglue 0.1cm

Let us recall some peculiarities of the multistep processes kinematics established
first in \cite{kop1,kop2} and described in details in \cite{long}. It is very selective 
kinematics, essentially different from the kinematics of the forward scattering off nuclei
when random walking of the particle is allowed in the plane perpendicular to the 
projectile momentum. Schematically the multistep
process is shown on Fig. 2.1.

{\bf Rescatterings.} 
For light particles (photon, also $\pi$-meson) iteration of the Compton formula
$${1\over \omega_{n} } -{1\over \omega_{n-1}} \simeq {1\over m} \left[1-cos (\theta_n)\right] \eqno (2.3)$$
allows to get the  final energy in the form                     
$$  {1\over \omega_N}- {1\over \omega_0} = {1\over m} \sum_{n=1}^N  \left[1-cos (\theta_n)\right] \eqno (2.4)  $$
The maximal energy of final particle is reached for the coplanar process when
all scattering processes take place in the same plane and each angle equals to
$\theta_k=\theta/N$.
As a result we obtain
$$  {1\over \omega_N^{max}}- {1\over \omega_0} = {1\over m} N \left[1-cos (\theta/N)\right] \eqno (2.5) $$
Already at $N>2$ and for $\theta \leq \pi$ the $1/N$ expansion can be made (it is in fact the $1/N^2$ expansion):
$$ 1-cos (\theta/N) \simeq \theta^2/2N^2 \left(1 - \theta^2/12N^2\right) \eqno (2.6)$$
and for large enough incident energy $\omega_0$ we obtain
$$\omega_N^{max} \simeq N {2m\over \theta^2} + {m\over 6N}. \eqno (2.7)$$
This expression works quite well beginning with $N=2$.
This means that the kinematically forbidden for interaction with single nucleon region
is partly filled up due to elastic rescatterings.
Remarkably, that this rather simple property of rescattering processes has not been
even mentioned in the pioneer papers \cite{baldin1} - \cite{leksin3}
\footnote{This property was well known, however, to V.M.Lobashev, who observed experimentally
that the energy of the photon after 2-fold interaction can be substantially
greater than the energy of the photon emitted at the same angle in
1-fold interaction.}.

In the case of the nucleon-nucleon scattering (scattering of particles with equal 
nonzero masses in general case) it is convenient to introduce the factor
$$ \zeta = {p\over E+m}, \qquad  1-\zeta^2 = {2m\over E+m},    \eqno (2.8)  $$
where $p$ and $E$ are spatial momentum and total energy of the particle
with the mass $m$. When scattering takes place on the particle which is at rest in
the laboratory frame, the $\zeta$ factor of scattered particle is multiplied
by $cos\,\theta$, where $\theta$ is the scattering angle in the laboratory frame.
So, after $N$ rescatterings we obtain the $\zeta$ factor
$$\zeta_N = \zeta_0 cos\,\theta_1 cos\,\theta_2 ... cos\,\theta_N. \eqno (2.9)$$
As in the case of the small mass of rescattered particle, the maximal value of final $\zeta_N$ is obtained when
all scattering angles are equal
$$ \theta_1 = \theta_2=...=\theta_N =\theta/N, \eqno (2.10)$$
and the coplanar process takes plase.  So, we have
$$\zeta_N^{max} = \zeta_0 \left[cos (\theta/N)\right]^N.\eqno (2.11) $$

The final momentum is from $(2.11)$
$$ k^{max} = 2m {\zeta ^{max}\over 1-\left(\zeta^{max}\right)^2} \eqno (2.12)$$
Again, at large enough $N$ and large incident energy ($\zeta_0 \to 1 $) the $1/N^2$ expansion can be made
at $k\gg m$, and we obtain the first terms of this expansion
$$k_N^{max} \simeq N {2m\over \theta^2} - {m \over 3N}, \eqno (2.13)$$
which coincides at large $N$ with previous result for the rescattering of light particles,
but preasymptotic corrections are negative in this case and twice greater \footnote{The preasymptotic
corrections given by Eqs. $(2.7)$ and $(2.13)$ are presented here for the first time}.

The normal Fermi motion of nucleons inside the nucleus makes these boundaries wider \cite{long}:
$$k_N^{max} \simeq N {2m\over \theta^2} \left[1 + {p_F^{max}\over 2m}\left(\theta + {1\over \theta} \right) \right], \eqno (2.14)$$
 where it is supposed that the final angle $\theta$ is large, $\theta \sim \pi$. For numerical estimates we
took the step function for the distribution in the Fermi momenta of nucleons inside
of nuclei, with $p_F^{max}/m\simeq 0.27$, see \cite{long} and references there. At large enough $N$
normal Fermi motion makes the kinematical boundaries for MIP wider by about $40$ \%.

There is characteristic decrease (down-fall) of the cumulative particle production cross
section due to simple rescatterings near the strictly backward direction. 
However, inelastic processes with excitations of intermediate
particles, i.e. with intermediate resonances, are able to fill up the region at $\theta \sim \pi$.

{\bf Resonance excitations in intermediate states.} 
The elastic rescatterings themselves are only the "top of the iceberg".
Excitations of the rescattered particles, i.e. production of resonances in intermediate
states which go over again into detected particles in subsequent interactions,
provide the dominant contribution to the production cross section.
Simplest examples of such processes may be $NN \to NN^* \to NN$, $\pi N\to \rho N \to \pi N$,
etc. The important role of resonances excitations in intermediate states for cumulative
particles production has been noted first by M.Braun and V.Vechernin \cite{bv-res-1} and somewhat later in
\cite{kop2}, see Figs. $(2.2)$ and $(2.3)$. At incident energy about few GeV the dominant contribution into cumulative protons
emission provide the processes with $\Delta(1232)$ excitation and reabsorption, see \cite{long} and
\cite{dakhno}. Experimentally the role of dynamical excitations in cumulative nucleons production
at intermediate energies has been extablished in \cite{komarov} and, at higher energy, in \cite{malki}.
 
When the particles in intermediate states are slightly 
excited above their ground states, approximate estimates can be made. Such resonances could be $\Delta 
(1232)$ isobar, or $N^*(1470), \; N^*(1520)$ etc. 
for nucleons, two-pion state or $\rho(770)$, etc for incident pions, $K^*(880)$ for kaons.
This case has been investigated previously with the result for the relative change (increase)
of the final momentum $k_f$ (Eq. $(8)$ of \cite{kop2})
$$ {\Delta k_f \over k_f} \simeq {1\over N} \sum_{l=1}^{N-1} {\Delta M_l^2\over  k_l^2}, \eqno (2.15) $$
or
$$ {\Delta k_f^2} \simeq {2\over N^3} \sum_{l=1}^{N-1} l^2\Delta M_l^2, \eqno (2.16) $$
with $\Delta M_l^2 = M_l^2 - \mu^2$, $k_l$ is the value of 3-momentum in the $l$-th intermediate state.
This effect can be explained easily: the additional energy stored in the mass of intermediate particle
is transfered to the kinetic energy of the final (cumulative) particle.

The number of different processes for the $N$-fold MIP is $(N_R+1)^{N-1}$, where $N_R$ is the number of
resonances making important contribution to the process of interest. The greatest kinematical advantage
has the process with resonance production at the $(N-1)$-th step of the whole process with subsequent
its deexcitation at the last step \footnote{In some of cascade calculations the important contribution
to the cumulative nucleons production gives the process with production of pions of not high energy
with its subsequent absorption by two-nucleon pair. This process can be, at least partly, to the processes
with resonance formation and reabsorption, because pions of moderate energies are produced mostly via
resonance formation and decay to nucleon and pion.}
. To calculate contributions of all these processe one needs not only to know 
 cross sections and the spin structure of the amplitudes $ N N^*_1 \to N N^*_2$ at the energies
up to several $Gev$, but also consider correctly possible interference betwee amplitudes of
different processes. Such information is absent and hardly will be available in nearest future.

To produce the final particle at the absolute boundary available for the nucleus as a whole 
one needs to have the masses of intermediate 
resonances (or some particles system)  of the order of incident energy, $s\sim E_0 m_A$.

\vglue 0.1cm
\begin{figure}[h]
\begin{center}

\includegraphics[natwidth=800,natheight=600,width=7.cm,angle=0]{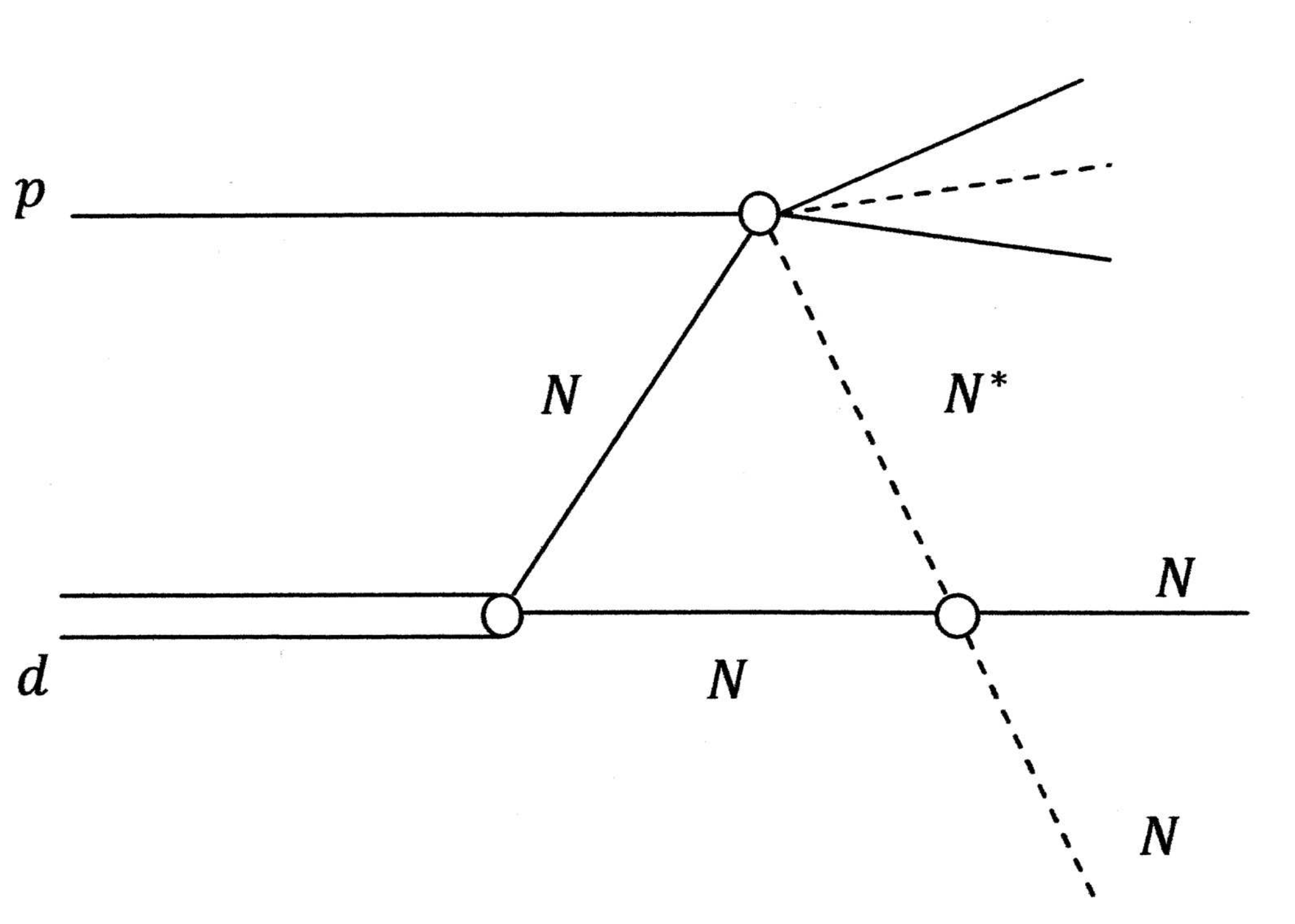}

\end{center}
{\small {\bf Fig.\ 2.2.} The diagram of the two-fold interaction process on the deuteron with the 
nucleon resonances (or $\Delta$ isobars) excitations in intermediate states.}
\end{figure}
\vglue 0.1cm

\vglue 0.1cm
\begin{figure}[h]
\begin{center}
\includegraphics[natwidth=800,natheight=600,width=9.cm,angle=0]{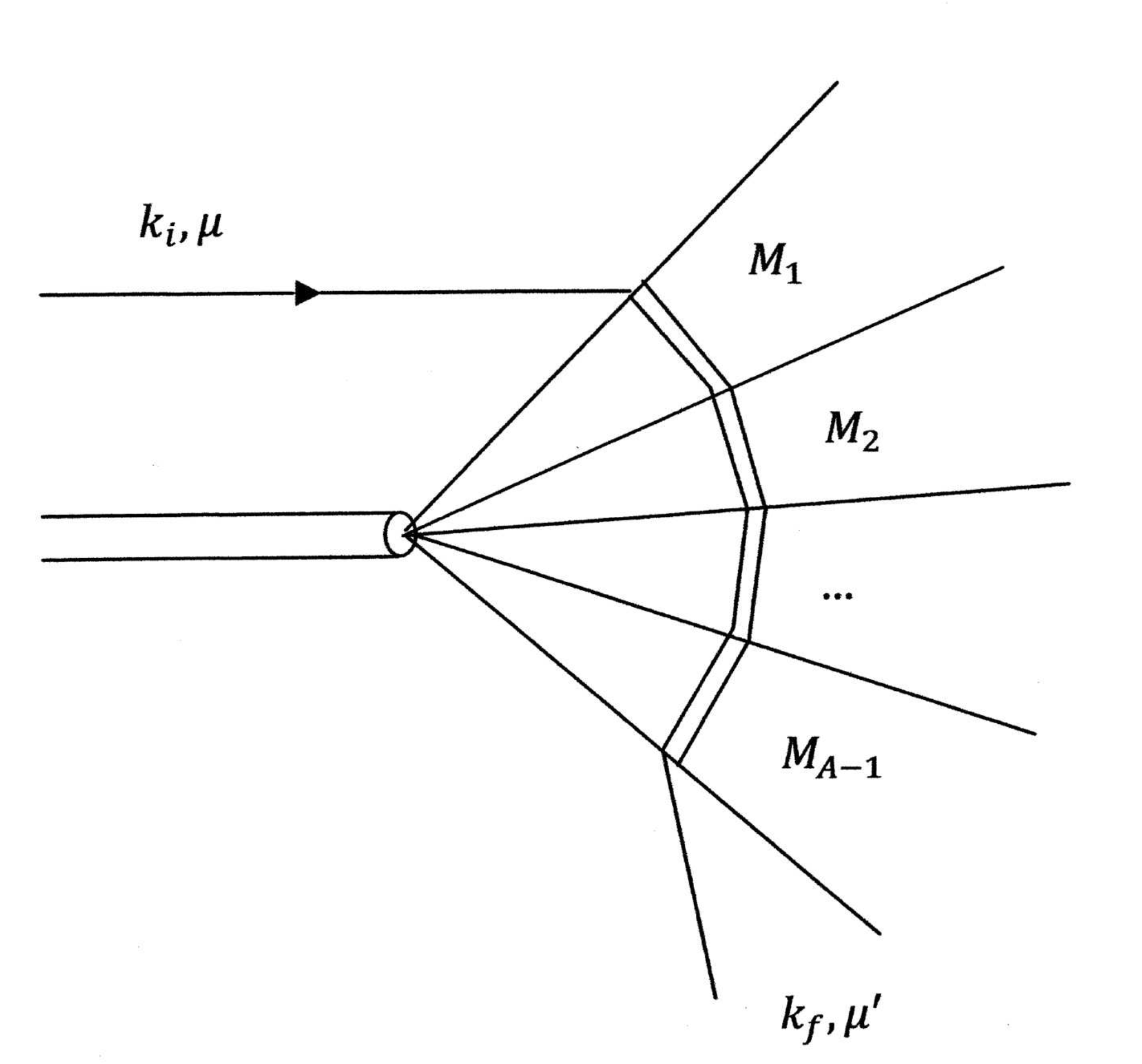}

\end{center}
{\small{\bf Fig.\ 2.3.} Schematical picture of the multiple interaction process with the resonances excitations
in intermediate states. The resonances may be either different or the same in different intermediate states.}
\end{figure}
\vglue 0.1cm

In this extreme case 
$$M_l^2(max) \simeq s_A {l\over A}\left(1 - {l\over A} \right) \eqno (2.17) $$
where $s_A \simeq 2AE_0m$ (\cite{kop2}, Appendix). Interaction with all $A$ nucleons should take place,
and the intermadiate mass is maximal at $l \sim A/2$.
For the deuteron the intermediate mass at the absolute boundary should be
$$M_1^D(max) \sim s_D/4 \simeq E_0 m/2. \eqno(2.18)$$

This case is of academic interest, only. Our aim is to show that the whole region of final particles momenta
allowed for interaction with the nucleus as a compact object can be covered due to MIP, but the price for this
are the extremely large masses of intermediate states.

What is the most important: at arbitrary high incident energy the kinematics of all
subsequent processes is defined by the momentum and the angle of the outgoing particle.
In other words, for the nucleus fragmentation with particles emitted backwards with probably large but
limited by few $GeV$ energies, the fragmentation of nucleon takes place in the first interaction act
of the MIP, according to kinematics analysed above. The slight dependence of the whole
MIP on the incident projectile, hadron or lepton, follows from this observation, as it was noted
long ago by Leksin et al \cite{leksin1}-\cite{leksin3}.

The theory of elementary particles based on the S-matrix approach operates with
so called $|in>$ and $<out|$ states as initial and final states of the process
under consideration. It is assumed that there is time enough for the formation
of the outgoing particles and the fields surrounding it.
 Usually it is in complete correspondence with
experimental conditions, when the elementary interaction amplitude is studied
by means of cross sections, polarization observables, etc. measurements.

Situation may be, however, quite different when the interaction of the projectile with
nucleons inside the nucleus takes place.
The role of the formation time in the interaction of the particle within some medium
has been discussed long ago, one of the pioneer paper is the paper by Landau and 
Pomeranchuk \cite{lapo} where the electromagnetic processes of the photon
emission and pair production by electrons has been considered.
Similar to the case of electromagnetic interactions, the hadron formation 
time is of the order of
$$\tau^{form} \sim 1/(\omega - k_z) \eqno (2.20)$$
if the incident energy is large enough,
where $\omega$ and $k_z$ are the energy and the longitudinal momentum of the
produced particle, the axis $z$ is defined by the momentum of the incident
particle.
When the particle is produced in the forward direction with large enough energy
(momentum), the formation time becomes
$$\tau^{form} \sim {2\omega \over \mu^2}, \eqno (2.21) $$
where $\mu$ is the mass of the produced particle.
So, formation time, or coherence length in forward direction, become very large for the 
energetic particle produced
in the direction of the projectile momentum (see, e.g. \cite{nnn} for review of the history of 
this problem and references. The nuclei fragmentation region has not been discussed in \cite{nnn}).

As noted above, for the production of a particle on a target with the mass $m_t$
at high enough incident energy the inequality takes place:
$$\omega - k_z \leq m_t, \eqno (2.22) $$
at the kinematical boundary the equality takes place.
As we have shown in this section, to produce a final particle beyond 
the kinematical boundary due to multiple interaction process, 
in the first interaction act the particle should be produced near the 
kinematical boundary, i.e.
$$\omega_1- cos \theta_1k_{1} \sim m_N, \eqno (2.23) $$
therefore, the formation time of the first produced particle 
$$\tau_1^{form} \sim 1/(\omega_1 - cos\theta_1 k_1) \sim 1/m \eqno (2.24)$$
is  necessarily small, and the whole production picture is
of quasiclassical character.
The interesting phenomena observed in the high energy particles - nuclei interaction 
reactions and widely discussed in the literature \cite{nnn}, 
connected with the large formation time of the particles produced 
in forward direction, do not take place in the cumulative production
processes.

\section{The small phase space method for the MIP probability calculations}
This method, most adequate for analytical and semi-analytical calculations of
the MIP probabilities, has been proposed in \cite{kop2} and developed later in 
\cite{long}. It is based on the fact that, according to established in \cite{kop1,kop2}
and presented in previous section kinematical relations, there is a preferable plane of the 
whole MIP leading to the production of energetic particle at large angle $\theta$, 
but not strictly backwards. Also, the angles of subsequent rescatterings are close
to $\theta / N$. Such kinematics has been called optimal, or basic kinematics.
The deviations of real angles from the optimal values are small, they are defined mostly
by the difference $k_N^{max} - k $, where $k_N^{max}(\theta)$ is the maximal possible momentum reachable
for definite MIP, and $k$ is the final momentum of the detected particle.
$k_N^{max}(\theta)$ should be calculated taking into account normal Fermi
motion of nucleons inside the nucleus, and also resonances excitation ---
deexcitation in the intermediate state. Some high power of the difference  $(k_N^{max} - k)/k_N^{max} $
enters the resulting probability.

Within the quasiclassical treatment adequate for our case, the probability product approximation is
valid, and the starting expression for the inclusive cross section of the particle
production at large angles contains the product of the elementary subprocesses matrix elements squared,
 see, e.g., Eq. (4.11) of \cite{long}.


After some evaluation, introducing differential cross sections of binary reactions $d\sigma_l/dt_l(s_l,t_l) $ instead of 
the matrix elements of binary reactions $M_l^2(s_l,t_l)$, we came to the formula for the production cross section 
due to the $N$-fold MIP \cite{kop2,long}
$$f_N(\vec p_0,\vec k)= \pi R_A^2 G_N(R_A,\theta) \int \frac{f_1(\vec p_0,\vec k_1) (k_1^0)^3 x_1^2dx_1 d\Omega_1}{\sigma_1^{leav}\omega_1}
\prod_{l=2}^N\left({d\sigma_l(s_l,t_l)\over dt_l}\right) \frac{(s_l-m^2-\mu_l^2)^2-4m^2\mu_l^2}{4\pi m\sigma_l^{leav} k_{l-1}} $$
$$\times \prod_{l=2}^{N-1}\frac{k_l^2 d\Omega_l}{k_l(m+\omega_{l-1}-z_l\omega_lk_{l-1})\;}{1\over \omega_N'}\delta(m+\omega_{N-1}-\omega_N-\omega_N').
\eqno(3.1) $$ 
Here $z_l= cos\, \theta_l$,
 $\sigma_l^{leav}$ is the cross section defining the removal (or leaving) of the rescattered object at the corresponding section
of the trajectory, it is smaller than corresponding total cross section. 
$G_N(R_A,\theta)$ is the geometrical factor which enters the probability of the $N$-fold multiple interaction with
definite trajectory of the interacting particles (resonances) inside the nucleus. This trajectory is defined mostly
by the final values of $\vec k$ $(k, \,\theta) $, according to the kinematical relations of previous section. 
Inclusive cross section of the rescattered particle production in the first interaction  is
$\omega_1 d^3\sigma_1/d^3k_1=f_1(\vec p_0,\vec k_1)$ and 
$d^3k_1=(k_1^0)^3x_1^2dx_1$,
$\omega_N=\omega$ --- the energy of the observed particle.

To estimate the value of the cross section $(3.1)$ one can extract the product of the cross sections out of the integral $(3.1)$ near the 
optimal kinematics and multiply by the small phase space avilable for the whole MIP under consideration \cite{kop1,long}.
Further details depend on the particular process. For the case of the light particle rescattering, $\pi$-meson for example, $\mu_l^2/m^2\ll 1$,
we have
$${1\over \omega_N'}\delta(m+\omega_{N-1}-\omega_N-\omega_N') = {1\over k k_{N-1}}
\delta\left[ {m\over k} - \sum_{l=2}^N (1-z_l) - {1\over x_1}\left({m\over p_0}+1-z_1\right)\right] \eqno(3.2) $$
To get this relation one should use the equality $\omega_N'=\sqrt{m^2+k^2+k_{N-1}^2-2k k_{N-1}z_N}$ for the recoil nucleon energy
and the well known rules for manipulations with the $\delta$-function.
When the final angle $\theta$ is considerably
different from $\pi$, there is a preferable plane near which the whole
multiple interaction process takes place, and only processes near this plane
contribute to the final output. At the angle $\theta =\pi$, strictly backwards,
there is azimuthal symmtry, and the processes from the whole interval of azimuthal 
angle $0< \phi < 2\pi$ provide contribution to the final output (azimuthal focusing, see next section).
A necessary step is to introduce azimuthal  deviations from this optimal
kinematics, $\varphi_k$, $k=1,\,...,N-1$; 
 $\varphi_N=0$ by definition of the plane of the process, $(\vec p_0, \vec k)$.
Polar deviations from the basic values, $\theta/N$, are denoted as $\vartheta_k$, obviously, 
 $\sum_{k=1}^N\vartheta_k = 0$. The direction of the momentum $\vec k_l$ after $l$-th 
interaction, $\vec n_l$,  is defined by the azimuthal angle $\varphi_l$ and the polar angle 
$\theta_l = (l\theta /N) +\vartheta_1+...+\vartheta_l$, $\theta_N=\theta $.

Then we obtain making the expansion in $\varphi_l$,  $\vartheta_l$ up to quadratic terms
in these variables:
$$z_k= (\vec n_k \vec n_{k-1}) \simeq cos (\theta/N) (1-\vartheta_k^2/2) -sin (\theta/N) \vartheta_k 
+ sin (k\theta/N) sin [(k-1)\theta/N] (\varphi_k -\varphi_{k-1})^2/2. \eqno (3.3)$$
In the case of the rescattering of light particles the sum enters the phase space of the process
$$ \sum_{k=1}^N (1- cos\vartheta_k) =  N[1-cos(\theta/N)] + cos(\theta/N)
\sum_{k=1}^N\bigg[-\varphi_k^2\, sin^2(k\theta/N) + $$
$$+{\varphi_k \varphi_{k-1}\over cos(\theta/N)}sin(k\theta/N)sin((k-1)\theta/N)\bigg] -{cos(\theta/N)\over 2} \sum_{k=1}^N \vartheta_k^2 \eqno (3.4)$$ 
To derive this equality we used that $\varphi_N=\varphi_0=0$ --- by definition of the plane of the MIP,
and the mentioned relation $\sum_{k=1}^N\vartheta_k = 0$.
We used also the identity,  valid for $\varphi_N=\varphi_0 =0$:
$${1\over 2}\sum_{k=1}^N \left(\varphi_k^2+\varphi_{k-1}^2 \right) sin(k\theta/N)sin[(k-1)\theta/N] = 
cos(\theta/N)\sum_{k=1}^N\varphi_k^2 sin^2(k\theta/N). \eqno (3.5)$$
It is possible to present the quadratic form in angular variables which enters $(3.4)$ in the canonical form and to perform integration easily, see
Appendix B and Eq. $(4.23)$ of \cite{long}, and also Appendix in present paper.
As a result, we have the integral over angular variables of the following form:
$$ I_N(\Delta_N^{ext}) = \int \delta\biggl[\Delta_N^{ext} - z_N^\theta \bigg(\sum_{k=1}^{N} \varphi_k^2 -\varphi_k\varphi_{k-1}/z_N^\theta
+\vartheta_k^2/2\biggr)\biggr] \prod_{l=1}^{N-1} d\varphi_ld\vartheta_l = $$
$$= \frac{\left(\Delta_N^{ext}\right)^{N-2} (\sqrt 2 \pi)^{N-1}}{J_N(z_N^\theta) \sqrt N (N-2)! \left(z_N^\theta \right)^{N-1}}, \eqno (3.6) $$
$ z_N^\theta = cos(\theta/N)$. Since the element of a solid angle $d\Omega_l= sin(\theta\,l/N)d\vartheta_l d\varphi_l $, we made here substitution
$sin(\theta\,l/N)\,d\varphi_l \to d\varphi_l$ and $d\Omega_l \to d\vartheta_l d\varphi_l $,
$z_N^\theta = cos(\theta/N)$. 
The whole phase space is defined by the quantity
$$\Delta_N^{ext}\simeq {m\over k} - {m\over p_0} -N(1 - z_N^\theta) -(1-x_1){m\over p_0} \eqno (3.7) $$
which depends on the effective distance of the final momentum (energy) from the kinematical boundary for the $N$-fold process.
The Jacobian of the azimuthal variables transformation squared is
$$ J_N^2(z) = Det\, ||a_N||, \eqno (3.8)$$
where the matrix $||a_N||$ defines the quadratic form $Q_N(z,\varphi_k)$ which enters the argument of the $\delta$-function in Eq. $(3.6)$:
$$ Q_N(z,\varphi_k) = a_{kl} \varphi_k \varphi_l =\sum_{k=1}^{N} \varphi_k^2 -{\varphi_k\varphi_{k-1}\over z}. \eqno (3.9)$$
For example,
$$ Q_3(z,\varphi_k) =  \varphi_1^2 + \varphi_2^2 -\varphi_1 \varphi_2/z; \quad Q_4(z,\varphi_k)= \varphi_1^2 +\varphi_2^2 +\varphi_3^2 -
(\varphi_1\varphi_2 + \varphi_2\varphi_3)/z ,\eqno (3.9a) $$
see next section and Appendix.

The phase space of the process in $(3.1)$ which depends strongly on $\Delta_N^{ext}$,  after integration over angular 
variables can be presented in the form
$$\Phi_N^{pions} = {1\over \omega_N'}\delta(m+\omega_{N-1}-\omega_N-\omega_N')\prod_{l=1}^Nd\Omega_l = {I_N(\Delta_N^{ext})\over k k_{N-1}}=
\frac{(\sqrt 2\pi)^{N-1}(\Delta_N^{ext})^{N-2}}{kk_{N-1}(N-2)!\sqrt N J_N(z_N^\theta)\left(z_N^\theta\right)^{N-1}} \eqno(3.10) $$

The normal Fermi motion of target nucleons inside of the nucleus increases the phase space considerably \cite{kop2, long}:
$$\Delta^{ext}_{N} = \Delta^{ext}_{N}|_{p_F=0} + \vec p^F_l\vec r_l/2m, \eqno (3.11) $$
where $\vec r_l =2m(\vec k_l-\vec k_{l-1})/k_lk_{l-1} $. A reasonable approximation is to take vectors $\vec r_l$ according to
the optimal kinematics for the whole process, and the Fermi momenta distribution of nucleons inside of the nucleus
in the form of the step function.
Integration over the Fermi motion leads to increase of the power of $\Delta_N^{ext}$ and change of numerical coefficients in
the expression for the phase space. Details can be found in \cite{kop2, long}, but they are not importanr for our 
mostly qualitative treatment here.

For the case of the nucleons rescattering there are some important differences from the light particle case, but
the quadratic form which enters the angular phase space of the process is essentially the same, with additional
coefficient:
$$\Phi_N^{nucleons} = 
{1\over k(m+\omega_{N-1})}\int \delta\left[\Delta^{ext}_{N,nucl} -\left(z_N^\theta\right)^N Q_N(\varphi_k) -{\left(z_N^\theta\right)^{N-2}\over 2}\sum_{l=1}^N\vartheta_l^2\right]
\prod_{l=1}^Nd\Omega_l =$$
$$=\left(\frac{\sqrt 2\pi}{\zeta_0 z^{N-1}}\right)^{N-1} \frac{(\Delta^{ext}_{N,nucl})^{N-2}}{(N-2)!\sqrt N J_N(z_N^\theta)}
\frac{(1-\zeta_N^2)(1-\zeta_{N-1}^2)}{4m^2\zeta_N} \eqno(3.12) $$
where
$$ \Delta^{ext}_{N,nucl} = \zeta_N-(1-x_1)\zeta_N{1-\zeta_1^2\over 1+\zeta_1^2} - {k\over m+\omega},  \eqno (3.13) $$
with $\zeta_N = \zeta_0 \left(z_N^\theta\right)^N,\; \zeta_1=\zeta_0z_N^\theta$.
As in the case of the light particle rescattering, the normal Fermi motion of nucleons inside the nucleus can be taken into account.

\section{The backward focusing effect (Buddha's light of cumulative particles)}
This is the sharp enhancement of the production cross section near the strictly 
backward direction, $\theta = \pi$. 
This effect has been noted first experimentally in Dubna (incident protons, final particles pions, protons and deuterons) \cite{baldin-77,foc-1}
and somewhat later by Leksin's group at ITEP (incident protons of 7.5 GeV/c, emitted protons
of 0.5 GeV/c) \cite{foc-2}.
This striking effect was not well studied previously, both experimentally
and theoretically.
In the papers \cite{kop2,long} where the small phase space method has been developed, 
it was noted that this effect can appear due to multiple interaction processes (see p.122 of \cite{long}). However, the
consideration of this effect was not detailed enough, the explicit angular dependence of the cross section near
backward direction, $\theta = \pi$, has not been established, estimates and comparison with data have not been made
\footnote{\small One of the authors (VBK) discussed the cumulative (backward) particles production off nuclei 
with professor Ya.A.Smorodinsky
who noted its analogy with known optical phenomenon - glory, or "Buddha's light". The
glory effect has been mentioned by Leksin and collaborators \cite{glo-1}, however,
it was not clear to authors of \cite{glo-1}, can it be related to cumulative production,
or not. In the case of the optical (atmospheric) glory phenomenon the light scatterings 
take place within droplets of
water, or another liquid. A variant of the atmospheric glory theory can be found in \cite{khare}. 
However, the optical glory is still not fully understood, the existing explanation 
is still incomplete, see, e.g. http://www.atoptics.co.uk/droplets/glofeat.htm. In nuclear physics
the glory-like phenomenon due to Coulomb interaction has been studied in \cite{maiorova} for the case of low energy
antiprotons (energy up to few KeV) interacting with heavy nuclei.}. 

The backward focusing effect has been observed and confirmed later in a number of
papers for different projectiles and incident energies \cite{foc-4,glo-1,glo-2}. 
It seems to be difficult to explain the backward focusing effect as coming from 
interaction with dense few nucleon clusters existing inside the nucleus.

Mathematically the focusing effect comes from the consideration of the small phase space of 
the whole multiple interaction process by the method described in previous section.
It takes place for any MIP, regardless the particular
kind of particles or resonances in the intermediate states.
As it was explained in section 2, when the angle of cumulative particle emission is large, but different from $\theta = \pi$,
there is a prefered plane for the whole process.
When the final angle $\theta = \pi$, then integration over one of azimuthal angles takes place
for the whole interval $[0, 2\pi]$, which leads to the rapid increase of the resulting cross section when the
final angle $\theta$ approaches $\pi$.. 

We show first that the azimuthal focusing takes place for any values of the polar scattering angles $\theta_k^{opt}$.
For arbitrary angles $\theta_k$ the cosine of the angle between directions $\vec n_k$ and $\vec n_{k-1}$ is
$$z_k= (\vec n_k \vec n_{k-1}) \simeq cos (\theta_k - \theta_{k- 1}) (1-\vartheta_k^2/2) -sin (\theta_k - \theta_{k- 1}) \vartheta_k 
+ sin (\theta_k) sin \theta_{k-1} (\varphi_k -\varphi_{k-1})^2/2. \eqno (4.1)$$

After substitution $sin \theta_k\varphi_k \to \varphi_k$ we obtain
$$z_k= (\vec n_k \vec n_{k-1}) \simeq cos (\theta_k - \theta_{k- 1}) (1-\vartheta_k^2/2) -sin ((\theta_k - \theta_{k- 1})) \vartheta_k 
+ {s_{k-1}\over 2s_k}\varphi_k^2 + {s_{k}\over 2s_{k-1}}\varphi_{k-1}^2-\varphi_{k-1})\varphi_k, \eqno (4.2)$$
where we introduced shorter notations $s_k = sin \theta_k$.

It follows from Eq. $(4.2)$ that tin general case of arbitrary polar angles $\theta_k$ the quadratic form depending on the 
small azimuthal deviations $\varphi_k$ which enters the sum $\sum_k (1-z_k)$ for the $N$-fold process is
$$Q_N^{gen}(\varphi_k,\varphi_l) = {s_2\over s_1}\varphi_1^2 + {s_1+s_3\over s_2}\varphi_2^2 +{s_2+s_4\over s_3}\varphi_3^2 +....
+ {s_{N-2}+s_N\over s_{N-1}}\varphi_{N-1}^2 -$$
$$-2 \varphi_1\varphi_2 -2\varphi_2\varphi_3 - ... - 2\varphi_{N-2}\varphi_{N-1} =||a||^{gen}(\theta_1,...,\theta_{N-1})_{kl}
\varphi_k\varphi_l , \eqno(4.3)$$
with $s_N= sin \theta$.
E.g., for $N=5$ we have the matrix
$$||a||_{N=5}^{gen}(\theta_1,\theta_2,\theta_3,\theta_4)=
\left[\begin {array}{cccc} s_{2}/s_{1} & -1 & 0 & 0\\ 
-1 & (s_{1}+s_{3})/s_{2} & -1  & 0   \\
 0 & -1  & (s_{2}+s_{4})/s_{3}& -1\\
0  &  0 & -1 & (s_{3}+s_{\theta})/s_{4}
   \end {array}\right], \eqno (4.4) $$
$s_\theta = s_5$, and generalization to arbitrary $N$ is straightforward.

Determinant of this matrix can be easily calculated.
It can be shown by induction that at arbitrary $N$
$$ Det \left(||a||_N^{gen}\right) = {s_\theta\over s_1}, \quad s_\theta = s_N. \eqno(4.5) $$

It follows from the generalized expression $(4.4)$ for the matrix $||a||$ that
$$Det ||a||_{N+1}^{gen} (\theta) = {s_{N-1}+s_\theta\over s_N} Det\left(||a||_N^{gen}\right)(\theta_N)- 
Det \left(||a||_{N-1}^{gen}\right)(\theta_{N-1}), \eqno (4.6) $$
where $\theta_{N+1}=\theta$. Since $Det \left(||a||_N^{gen}\right)(\theta_N) = {s_N/ s_1} $ and 
$Det \left(||a||_N^{gen}\right)(\theta_{N-1}) = {s_{N-1}/ s_1} $,  we obtain easily
$$Det ||a||_{N+1}^{gen} (\theta) = \left({s_{N-1}+s_\theta\over s_N}\right) {s_N\over s_1} - {s_{N-1}\over s_1} = {s_\theta\over s_1}. \eqno (4.7) $$

After integration the delta-function containing the quadratic form over the small azimuthal deviations we obtain
$$ \int \delta \left(\Delta - ||a||^{gen}_N(\theta_1,...,\theta_{N-1})_{kl}\varphi_k\varphi_l \right) d\varphi_1... d\varphi_{N-1} 
= {\Delta^{(N-3)/2}\over Det ||a||_N^{gen} (N-3)!!} (2\pi)^{(N-3)/2}c_{N-3}= $$
$$=\sqrt{{s_1\over s_\theta}} {\Delta^{(N-3)/2}\over  (N-3)!!} (2\pi)^{(N-3)/2}c_{N-3} , \eqno(4.8) $$
$c_n=\pi$ for odd $n$, and $c_n=\sqrt{2\pi}$ for even $n$, and $N-3\geq 0$, see Appendix.

We obtain from above expressions the characteristic angular dependence of the cumulative particles production cross section  near $\theta =\pi$:
$$ d\sigma \sim \sqrt{{s_1\over s_\theta}} \simeq \sqrt{{s_1\over \pi-\theta}}, \eqno(4.9) $$
since $sin\theta \simeq \pi -\theta$ for $\pi-\theta \ll 1$.

This formula does not work at $\theta = \pi$, because
integration over the azimuthal angle which defines the plane of the whole MIP takes place in the interval $(0, 2\pi)$. The result for the 
cross section is final, of course, as we show in details for the case of the optimal kinematics.

For the optimal kinematics with equal polar scattering angles $\theta_k= k\theta/N$ (see section 2), and the general quadratic form 
goes over into quadratic form obtained in \cite{long} with some coefficiens:
$$ Q^{gen} \to  2 z_N^\theta Q(z_N^\theta, \varphi_k,\varphi_l), \quad z_N^\theta = cos(\theta/N), \eqno (4.10) $$
and
$$Det (||a||_N^{gen}) =\left(2 z^\theta_N\right)^{N-1} Det (||a||_N).  \eqno (4.10a) $$
It is convenient to present the quadratic form which enters the $\delta$ - function in $(3.6)$ as
$$ Q_N(z_N^\theta,\varphi_k,\varphi_l) =J_2^2\left(\varphi_1-{\varphi_2\over 2zJ_2^2}\right)^2 +{J_3^2\over J^2_2}
\left(\varphi_2-{J_2^2\varphi_3\over 2zJ_3^2}\right)^2 +...$$
$$...  + {J_{N-1}^2\over J_{N-2}^2}\left(\varphi_{N-2}-{J_{N-2}^2\varphi_{N-1}\over 2z J_{N-1}^2}\right)^2 +
{J_N^2\over J_{N-1}^2} \varphi_{N-1}^2 .\eqno (4.11)$$
For the sake of brevity we omitted here the dependence of all $J_k^2$ on their common argument $z_N^\theta $.  The recurrent relation 
$$J_N^2(z)= J_{N-1}^2(z)-{1\over 4z^2}J_{N-2}^2(z) \eqno (4.12) $$ 
can be obtained from $(4.11)$, since, as it follows from$(3.6)$ and $(3.9)$
$$Q_{N+1}(z,\varphi_k,\varphi_l)  = Q_N(z, \varphi_k,\varphi_l) +\varphi_N^2 -\varphi_N\,\varphi_{N-l}/z \eqno (4.13)$$
(recall that for the $N+1$-fold process $\varphi_{N+1} =0$ by definition of the whole plane of the process),
The proof of relation $(4.12)$ is given in Appendix.

The following formula for $J_N^2(z_N^\theta)$ has been obtained in \cite{long}:
$$Det ||a_{kl}|| =J_N^2(z_N^\theta) = 1 + \sum_{m=1}^{m < N/2}\left(-{1\over 4\left(z_N^\theta\right)^2}\right)^m 
{\prod_{k=1}^{m}(N-m-k)\over m!} = $$
$$= 1 + \sum_{m=1}^{m < N/2}\left(-{1\over 4\left(z_N^\theta\right)^2}\right)^m C^m_{N-m-1}. \eqno (4.14)$$

Recurrent relations for Jacobians with subsequent values of $N$ and with same argument $z$:
$$ J^2_{N+1}(z) = J^2_N(z) -{1\over 4z^2} J^2_{N-1}(z)= 
J^2_{N-1}(z)\left(1-{1\over 4z^2}\right) -{1\over 4z^2} J^2_{N-2}(z)\eqno (4.15) $$
can be continued easily to lower values of $N$ and also used for calculations of $J_N^2$ at any $N$ starting
from two known values, $J_2^2(z) =1$ and $J_3^2(z) =1-1/(4z^2)$  (see Appendix).
The Eq. $(4.14)$ can be confirmed in this way.

The condition $J_N(\pi/N) =0$ leads to the equation for $z_N^\pi$ which solution
(one of all possible roots) provides the value of $cos(\pi/N)$ in terms of radicals.
The following expressions for these jacobians take place \cite{kop2,long}
$$J_2^2(z)=1; \qquad J_3^2(z) = 1-{1\over 4z^2}; \qquad J_4^2(z)= 1-{1\over 2z^2}, \eqno (4.16) $$
$J_3(\pi/3)=J_3(z=1/2)=0$,   $J_4(\pi/4)=J_4(z=1/\sqrt 2)=0$.
Let us give here less trivial examples. For $N=5$ 
$$J_5^2=1-{3\over 4z^2}+{1\over 16 z^4}, \qquad  (J_5^2)'_z = {3\over 2z^3} - {1 \over 4 z^5} \eqno (4.17) $$
and one obtains $cos^2(\pi/5) = (3+\sqrt 5)/8$, $J_5(\pi/5)=0$.

At $N=6$
$$J_6^2=1-{1\over z^2}+{3\over 16 z^4} = J_3^2 \left(1-{3\over 4z^2}\right), \qquad (J_6^2)'_z = {2\over z^3} - {3\over 4 z^5} . \eqno (4.18)$$
see also Eq. (A.9).
For $N=7$
$$J_7^2=1-{5\over 4z^2}+{3\over 8z^4} -{1\over 64z^6}, \qquad  (J_7^2)'_z = {5\over 2 z^3} - {3\over 2 z^5} + {3\over 32 z^7}. \eqno (4.19)$$
$J_7(\pi/7)=0$.

$$J_8^2=1-{3\over 2z^2}+{5\over 8z^4} -{1\over 16z^6}=J_4^2 \left(1-{1\over z^2}+{1\over 8z^4}\right), 
\qquad (J_8^2)'_z = {3\over z^3} - {5\over 2 z^5} +{3\over 8 z^7}, \eqno (4.20)$$
see Eq. (A.9);   $J_8(\pi/8)=0$.
For arbitrary $N$,  $J_N^2$ is a polinomial in $1/4z^2$ of the power $|(N-1)/2|$ (integer part of $(N-1)/2$), see Eq. $(4.14)$.
These equations can be obtained using the elementary mathematics methods as well, see Appendix, Eqs $(A.14)-(A.16)$.
The case $N=2$ is a special one, because $J_2(z)=1$ - is a constant. In this case the 2-fold
process at $\theta =\pi$ (strictly backwards) has no advantage in comparison with the direct one, see Eq. $(2.5)$,
if we consider the elastic rescatterings.

For particles emitted strictly backwards the phase space has different form, instead of
$J_N(\theta/N)$ enters $J_{N-1}(\theta/N)$ which is different from zero at $\theta =\pi$, and we have instead of Eq. (3.6)
$$ I_N(\varphi, \vartheta) = \int \delta\biggl[\Delta_N^{ext} - z_N^\pi\bigg(\sum_{k=1}^{N} \varphi_k^2 -\varphi_k\varphi_{k-1}/z_N^\pi
+\vartheta_k^2/2\biggr)\biggr] \left[\prod_{l=1}^{N-2} d\varphi_ld\vartheta_l\right] 2\pi d\vartheta_{N-1} = $$
$$= \frac{\left(\Delta_N^{ext}\right)^{N-5/2} (2\sqrt 2 \pi)^{N-1}}{J_{N-1}(z_N^\pi) \sqrt N (2N-5)!! \left(z_N^\pi\right)^{N-3/2}}, \eqno (4.21) $$
This follows from Eq. $(4.11)$ where at $\theta = \pi$ the last term disappears, since $J_N(\pi/N) = 0$ and integration over
$d \varphi_{N-1}$ takes place over the whole $2\pi$ interval.

\vglue 0.1cm
\begin{figure}[h]
\begin{center}
\includegraphics[natwidth=800,natheight=600,width=13.cm,angle=0]{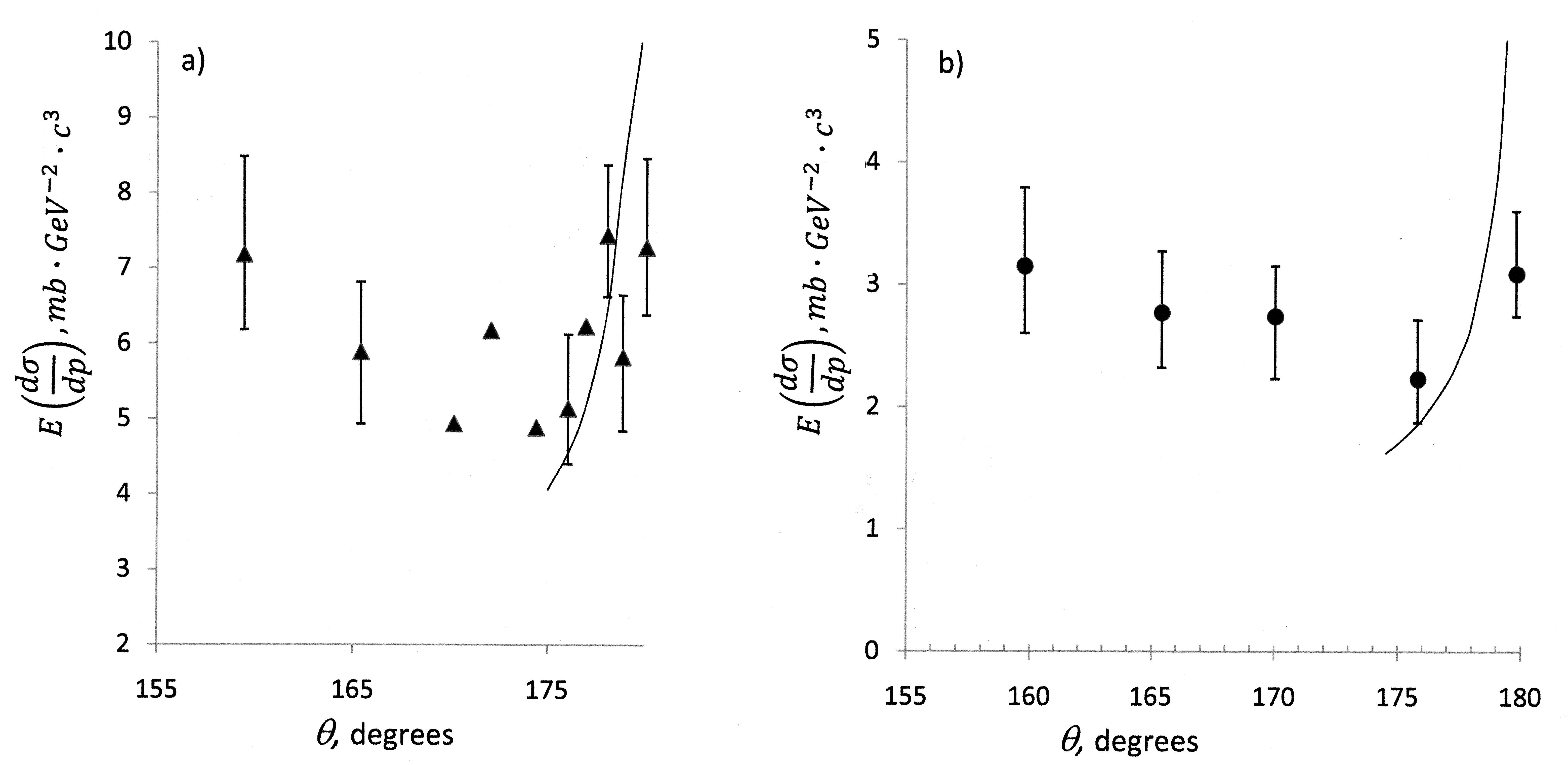}

\end{center}
{\small{\bf Fig.\ 4.1.} The angular dependence of inclusive cross section of the production of positive pions
by projectile protons with momentum $8.9\, GeV/c$. a) pions with momentum $0.5\, GeV/c$ emitted from  $Pb$ nucleus.
The error bars at some points have not been clearly indicated in the original paper; 
b) pions with  momentum $0.3\, Gev/c$ emitted from  $He$ nucleus.
 The data are taken from Fig. 18 of the paper \cite{foc-1}.  }
\end{figure}

\begin{figure}[h]
\begin{center}
\includegraphics[natwidth=800,natheight=600,width=15.cm,angle=0]{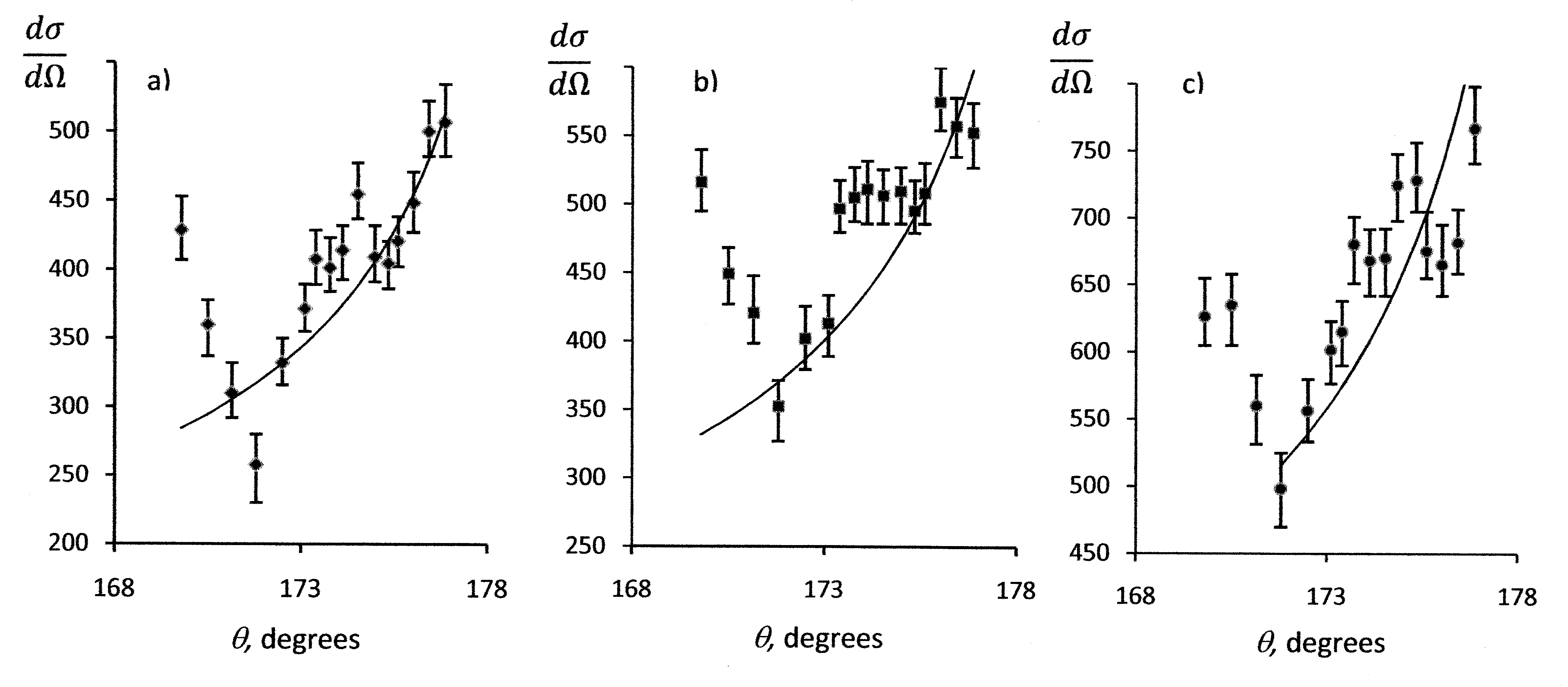}

\end{center}
{\small {\bf Fig.\ 4.2.} Angular distributions of secondary protons with kinetic energy between $0.06$ and $0.24\, GeV$ 
emitted from the Pb nucleus, in arbitrary units. The momentum of the projectile protons is $4.5\, GeV/c$. 
 a) The energy of emitted protons in the interval $0.11 \, - \, 0.24\, GeV$; b) the energy interval $0.08\,- \, 0.11\, GeV$; 
c) the energy interval  $0.06\,-\, 0.08\, GeV$.  Data obtained by G.A.Leksin group at ITEP, taken 
from Fig. 3 of paper \cite{glo-1}.}
\end{figure}
\vglue 0.1cm

To illustrate the azimuthal focusing which takes place near $\theta = \pi$ the ratio is useful of 
the phase spaces near the backward direction and  strictly at $\theta =\pi$. The ratio of the observed cross sections in the interval
of several degrees slightly depends on the elementary cross sections and is defined mainly by this ratio of phase spaces.
It is
$$ R_N(\theta) = {\Phi(z)\over \Phi (\theta =\pi)} = \sqrt{\Delta_N^{ext}\over z_N^\pi}  {(2n-5)!! \over 2^{N-1} (N-2)!}{J_{N-1}(z_N^\pi)\over sin(\pi/N) J_N(z_N^\theta)} 
\eqno (4.22)  $$
Near $\theta = \pi$ we use that 
$$J_N(z_N^\theta) \simeq \sqrt{{\pi - \theta\over N} [J_N^2]'(z_N^\pi) sin{\pi\over N} } \eqno (4.23) $$
and thus we get
$$ R_N(\theta) = C_N \sqrt{{\Delta_N^{ext}\over \pi - \theta}} \eqno (4.24) $$
with
$$ C_N = {J_{N-1}(z_N^\pi) \sqrt N \over [(J_N^2)'(z_N^\pi)]^{1/2} [sin(\pi/N)]^{3/2}} { (2N-5)!! \over \sqrt{z_N^\pi} (N-2)! 2^{N-1}} \eqno(4.25)  $$
We need also values of $J_{N-1}[\pi/N] $ to estimate the behaviour of the cross sction near $\theta =\pi$,
they are given in Table 1.
Integration over variable $x_1$ leads to multiplication $C_N$ by factor $(2N-3)/(2N-2)$, i.e. it makes it smaller, increasing the
effect under consideration.

According to Eq. $(4.15)$, the differential cross section of the cumulative particle production increases with increasing
angle $\theta$. 
At the critical value 
$$ \theta^{crit}\simeq \pi - C_N^2 \Delta^{ext}_N, \quad \epsilon^{crit}=\pi -\theta^{crit} \simeq C_N^2 \Delta ^{ext} \eqno(4.26) $$
it becomes equal to the cross section at $\theta = \pi$ which is proportional to Eq. $(4.12)$, and may slightly increase further
with increasing $\theta$. But near $\theta = \pi$ it should derease, to become again $d\sigma /d\Omega|_{\theta=\pi}$
at $\theta=\pi$.  So, the differential cross section has a crater-like (or funnel-like) form near the backward direction.
We do not provide here the detailed description of the cross section in the transition region between $\theta^{crit}$ and $\theta=\pi$:
this is technically rather complicated problem, and not so important for us now.
\begin{center}
\begin{tabular}{|l|l|l|l|l|l|l|l|}                   
\hline
$N$& $(J_N^2(z_N^\pi))'$& $ sin(\pi/N)$ & $\left[(J_N^2(z_N^\pi))'\,sin^3 (\pi/N)\right]^{1/2} $& $J_{N-1}[z_N^\pi]$&$C_N $\\
\hline
$3$ & $\;\,4$ & $0.866 $&$\qquad  1.612$ &$\;\,1$ & $0.38 $\\
\hline
4& 2.83 & 0.707&\qquad 0.999 &0.707 & 0.32\\
\hline
5& 2.11 & 0.588 &\qquad  0.655 &0.486& 0.29\\
\hline
6& 1.540 & 0.5 &\qquad  0.438 & 0.333& 0.27\\
\hline
7& 1.087 & 0.434 & \qquad 0.298 &0.229& 0.26 \\
\hline
\end{tabular}
\end{center}
{\small {\bf Table 1.} Numerical values of the quantities which enter the particles production cross section
near backward direction, $\theta = \pi$}. Here $z_N^\pi=cos(\pi/N)$.\\

Characteristic values of $\Delta^{ext}$ are defined by kinematical boundaries described in section 2, Eq. $(2.7),\,(2.13)$, and
we obtain easily
$$\Delta^{ext}_{typical} \sim \theta^2/2N(N+1) < \pi^2/2[N(N+1)], \eqno(4.27) $$
so it is not greater than $\sim 0.5$ for $N=3$ and decreases rapidly with increasing $N$. Therefore, the values of $\epsilon^{crit}$
may be quite small, about several degrees.

Inclusion of resonance excitation in one (or several) intermediate states leads to the increase of the quantity $\Delta^{ext}_N$
according to formulas of section 2, and to the increase of the phase space of the whole MIP, but the effect of azimuthal focusing
persists.
Quite similar results can be obtained for the case of nucleons, only some technical detaols are different, see section 3.
The inclusion of the normal Fermi motion of nucleons inside the nucleus increases the
values of $\Delta_N^{ext}$, but numerical coefficient in $C_N$ becomes smaller.
The behaviour given by Eq. (4.15) is in good agreement with available data, the value of the constants $C_N$ is not
important for our semiquantiatative treatment.
The comparison of the observed behaviour with predicted one according to the simple law $d\sigma \sim A + B/\sqrt{\pi - \theta}$
is presented in Fig.4.1,  Fig. 4.2 and Fig. 4.3. 
\begin{figure}[h]
\begin{center}

\includegraphics[natwidth=800,natheight=600,width=5.5cm,angle=0]{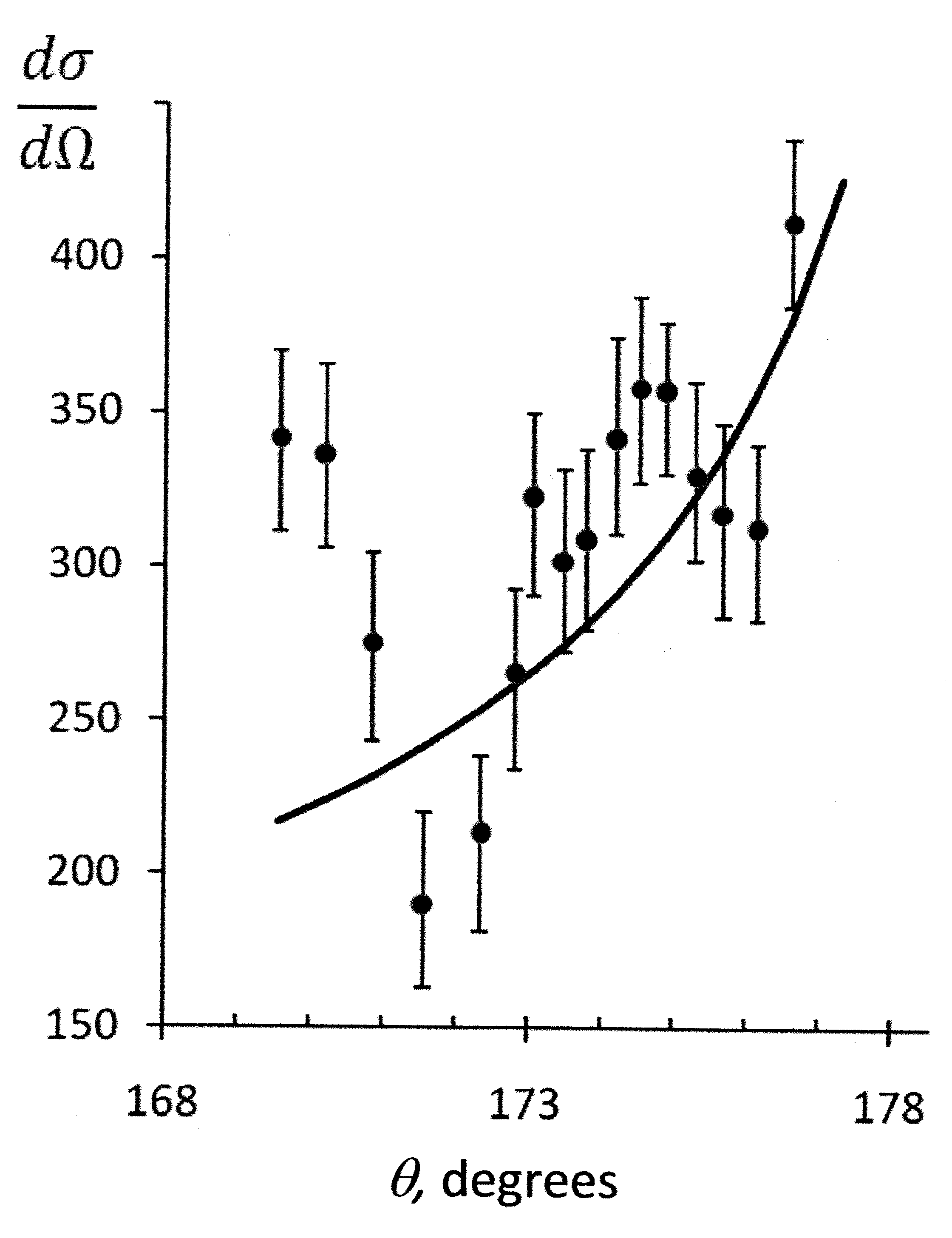}
\end{center}
{\small {\bf Fig.\ 4.3.} Angular distributions of secondary pions with kinetic energy greater $0.14\, GeV$ 
emitted from the Pb nucleus, in arbitrary units. The momentum of the projectile protons is $4.5$ GeV/c. 
  Data obtained by G.A.Leksin group at ITEP, taken from Fig. 5 of paper \cite{glo-1}. }
\end{figure}

We selected several examples where qualitative agreement
of data with predicted behaviour takes place. 
In Fig. 4.1 the inclusive cross section of the production of positive pions
by projectile protons with momentum $8.9\, GeV/c$ is presented for pions with momentum $0.5\, GeV/c$ ( $Pb$ as a target)
and for pions with  momentum $0.3\, Gev/c$ ($He$ as a target)\cite{foc-1}.
In Fig 4.2 angular distributions of secondary protons with kinetic energy between $0.06$ and $0.24\, GeV$ 
emitted from the Pb nucleus are presented, in arbitrary units. The momentum of the projectile protons is $4.5\, GeV/c$ \cite{glo-1}.
In Fig 4.3 angular distributions of secondary pions with kinetic energy greater $0.14\, GeV$ 
emitted from the Pb nucleus, are presented, also in arbitrary units. The momentum of the projectile protons is $4.5$ GeV/c. 
  Data are taken from Fig. 5 of paper \cite{glo-1}.

There are other data where the glory-like effect is clearly seen.
In many other cases the flat behaviour of the differential cross section
near $\theta \sim \pi$ takes place, but it was probably not sufficient resolution to detect the enhancement
of the cross section near $\theta = \pi$.
In some experiments  the deviation
of the final angle from 180 deg. is large, therefore, further measurements near $\theta = \pi$ are desirable,
also for kaons, hyperons as cumulative particles.
\section{Discussion and conclusions}
The nature of the cumulative particles is complicated and not well understood so far.
There are different possible sources of their origin, including the color forces \cite{kopnied}, one of them are the multiple 
collisions inside the nucleus, i.e. elastic or inelastic rescatterings.
We have shown that the enhancement of the particles production cross section off nuclei near
the backward direction, the glory-like backward focusing effect, is a natural property
of the multiple interaction mechanism for the cumulative particles production.
It takes place for any multiplicity of the process, $N\geq 3$, when the momentum of the emitted particle is close 
to the corresponding kinematical boundary. The universal dependence of the cross section,
$d\sigma \sim 1/\sqrt{\pi - \theta}$ near the final angle $\theta \sim \pi$, takes place regardless 
the multiplicity of the process. This statement by itself is quite rigorous and presented for the first time in the literature. 
The competition of the
processes of different multiplicities can make this effect difficult for observation in some cases.
Presently we can speak only about qualitative,
in some cases semiquantitative agreement with data. It is not clear yet how the transition to 
strictly backward direction proceeds. The angular distribution of emitted
particles near $\theta = \pi$ can have a narrow dip, i.e. it may be of a crater (funnel)-like form. Further studies are 
necessary for better understanding.

This effect, observed in a number of experiments at JINR and ITEP, is a clear manifestions 
of the fact that multiple interactions make important contribution
to the cumulative particles production probability, although it does not exclude the contribution
of interaction of the projectile with few-nucleon, or multiquark clusters possibly existing in nuclei.
We have proved the existence of the azimuthal focusing for arbitrary polar angles (rescattering of the light particles) and for the 
case of the optimal (basic) configuration of the MIP, also for nucleons rescattering. Investigation of other possible variants of the
optimal kinematical configurations, besides those considered in present paper may be of interest, but obviously, the azimuthal
focusing, discussed e.g. in \cite{khare} for the optical glory phenomenon, takes place for any kind of
MIP; only some technical details are different.

It would be important to detect the focusing effect for different types of produced
particles, baryons and mesons. This effect can be considered as a "smoking gun" of the MIP mechanism. 
If this nuclear glory-like phenomenon is observed for all kinds
of cumulative particles, its universality would be a strong argument in favor of importance of MIP.
Reactions where such effect is not observed would provide more chances for revealing nontrivial peculiarities 
of nuclear structure.
The role of the multiple interaction processes leading to the large
angle particles production off nuclei is certainly underestimated, still, by many authors,
theoreticians and experimentalists. Further efforts are necessary to settle this extremely difficult and important 
challenge of disentangling between the nontrivial effects of the nuclear structure and the MIP contributions.
\section{ Acknowledgements}
We are indebted to academician V.M.Lobashev who supported strongly 
the main idea that the background
multiple interaction processes should be investigated and their contribution
should be subtracted from measured cross sections to determine the weight of few-nucleon or 
multiquark clusters in nuclei.

We are thankful to Stepan Shimansky, whose questions, remarks and activity
stimulated appearance of present paper, and also to
A.B.Kurepin, V.L.Matushko for useful discussions. We thank Boris Kopeliovich, Anna Krutenkova and also referees for careful reading 
the manuscript and many useful remarks and suggestions.

The work is supported in part by Fondecyt (Chile), grant number 1130549.

\section{Appendix}
Here we present for the readers convenience some formulas and relations which have been used in sections 3 and 4.
$$I_n(\Delta)=\int \delta(\Delta - x_1^2-... - x_n^2) dx_1... dx_n = \pi {(2\pi)^{(n-2)/2}\over (n-2)!!} \Delta^{(n-2)/2} \eqno(A.1)$$
for integer even $n$.

$$I_n(\Delta)_n= \int \delta(\Delta - x_1^2-... - x_n^2) dx_1... dx_n =  {(2\pi)^{(n-1)/2}\over (n-2)!!} \Delta^{(n-2)/2} \eqno(A.2)$$
for integer odd $n$.
Relations
$$ \int_0^\pi sin^{2m}\theta\;d\theta\,=\,\pi{(2m-1)!!\over (2m)!!}; \qquad \int_0^\pi sin^{2m-1}\theta\;d\theta\,=\, 2{(2m-2)!!\over (2m-1)!!}, \eqno(A3) $$
$m$ --- integer, allow to check $(A1)$ and $(A2)$ easily.

The equality takes place
$$\int \delta(\Delta - x_1^2-... - x_n^2)\delta(x_1+x_2+... +x_n) dx_1... dx_{n-1} dx_n= 
 {1\over \sqrt n}  I_{n-1}(\Delta)  \eqno(A.4)$$
More generally, for any quatratic form in variables $x_k, \; k=1, ... n$ after diagonalization we obtain  
$$ \int \delta(\Delta - a_{kl}x_kx_l)dx_1\, ... \, dx_n= 
\int \delta(\Delta - x_1'^2-... - x_n'^2) {dx_1'...dx_n' \over \sqrt{det ||a||}} =
{1 \over \sqrt{det ||a||}}I_n(\Delta). \eqno(A.5) $$

Let $t$ be the transformation (matrix) which brings our quadratic form to the canonical form:
$$ \tilde t\, a\, t = {\cal I}, \eqno(A.6)$$
where ${\cal I}$ is the unit matrix $n\times n$, and $\tilde t_{kl} = t_{lk}$.
Then the equality takes place for the Jacobian of this transformation
$$ (det\,||t||)^{-2} = J_a^2(z) = det\,||a||, \qquad (det\,||t||)^{-1} = J_a(z) = \sqrt{det\,||a||}. \eqno(A.7)$$

To obtain the relation $(4.12)$ we write first the recurrent relation for the quadratic form
$$Q_{N+1}(z,\varphi_k,\varphi_l)  = Q_N(\varphi_k,\varphi_l) +\varphi_N^2 -\varphi_N\,\varphi_{N-l}/z, \eqno (A.8)$$
then rewrite it similar to Eq. $(4.11)$ and write down the equality for the last several terms
$$ {J_N^2\over J_{N-1}^2} \varphi_{N-1}^2 + \varphi_N^2 -{\varphi_N \varphi_{N-1}\over z} =
{J_N^2\over J_{N-1}^2} \left(\varphi_{N-1} - {J_{N-1}^2\over J_{N}^2}{\varphi_N \over 2 z}\right)^2 + {J_{N+1}^2\over J_{N}^2}\varphi_N^2. \eqno(A.9)$$
From equality of coefficients before $\varphi_N^2$ in the left and right sides we obtain
$$ 1 = {J_{N-1}^2\over 4z^2 J_{N}^2} + {J_{N+1}^2\over J_{N}^2} \eqno(A10), $$
and equation $(4.12)$ follows immediately.

The relation can be obtained from Eq. (4,12)
$$J_N^2(z) = J_{N-k}^2(z)J_{k+1}^2(z)-{1\over 4z^2}J_{N-k-1}^2(z)J_k^2(z) \eqno(A.11)$$
which, at $N=2m,\;k=m$ ($m$ is the integer),  leads to remarkable relation
$$ J_{2m}^2(z)= J_m^2(z)\left(J_{m+1}^2(z) -\,{1\over 4z^2} J_{m-1}^2(z) \right). \eqno(A.12) $$
Relation $(A.10)$ can be verified easily for $J_4^2,\;J_6^2$ and $J_8^2$,  see section 4.
It follows from $(A.10)$ that at $N=2m$ not only $J_N(\pi/N) =0$, but also
$J_N(2\pi/N)=0$ which has quite simple explanation.

For the odd values of $N$ another useful factorization property takes place:
$$ J_{2m+1}^2(z)= \left(J_{m+1}^2(z)\right)^2-{1\over 4z^2}\left(J_{m}^2(z)\right)^2 =\left(J^2_{m+1}(z) -{1\over 2z} J_{m}^2(z) \right)  
\left(J^2_{m+1}(z) +{1\over 2z} J_m^2(z) \right), \eqno(A.13) $$
which can be easily verified for $J_7^2$ and $J_5^2$ given in section 4.

The polinomials $J_N^2$ and equations for $z_N^\pi=cos(\pi/N)$ can be obtained in more conventional way.
There is an obvious equality
$$ [exp (i\pi/N)]^N = exp(i\pi) = -1 \eqno(A.14)$$
It can be written in the form
$$[cos(\pi/N) +i sin (\pi/N)]^N = -1, \eqno(A.15)$$
or separately for the real and imaginary parts
$$Re \left\{ [cos(\pi/N) +i sin (\pi/N)]^N\right\} = -1, \qquad Im \left\{ [cos(\pi/N) +i sin (\pi/N)]^N\right\} = 0. \eqno(A.16)$$
The polinomials in $z_N^\pi=cos(\pi/N)$ which are obtained in the left side of $(A.13)$ coincide with polinomials obtained in section 4.
However, some further efforts are necessary to get recurrent relations $(A.9),\,(A.10)$.

The following useful relations can be verified:
$$ \left(2z_N^\theta\right)^{N-1} J_N^2\left(z_N^\theta\right)sin{\theta\over N} = sin \theta. \eqno(A.17) $$
These relations provide the link between the general case considered at the beginning of section 4 and the particular 
case of the optimal kinematics with all scattering angles equal to $\theta/N$.

\vglue 0.2cm

\end{document}